\documentclass[a4paper]{article}

\usepackage{icrc2013}
\usepackage{amsmath}

\title{Influence of aerosols from biomass burning on the spectral analysis of Cherenkov telescopes}

\shorttitle{Influence of aerosols on gamma-ray spectral analysis}

\authors{
R. de los Reyes$^{1}$,
J. Hahn$^{1}$,
K. Bernl\"ohr$^{1}$,
P. Kr\"uger$^{1,2}$,
C. Deil$^{1}$,
H. Gast$^{3}$,
K. Kosack$^{4}$,
V. Marandon$^{1}$.
}

\afiliations{
$^1$ Max-Planck-Institut f\"ur Kernphysik, P.O. Box 103980, D 69029 \\
$^2$ North-West University, Potchefstroom, South-Africa \\
$^3$ RWTH Aachen University, Germany \\
$^{4}$ CEA Saclay, France
}

\email{reyes@mpi-hd.mpg.de}

\abstract{During the last decade, imaging atmospheric Cherenkov telescopes (IACTs) have proven themselves as astronomical detectors in the very-high-energy (VHE; E$>$0.1 TeV) regime.
The IACT technique observes the VHE photons indirectly, using the Earth's atmosphere as a calorimeter. Much of the calibration of Cherenkov telescope experiments is done using Monte Carlo simulations of the air shower development, Cherenkov radiation and detector, assuming certain models for the atmospheric conditions. 
Any deviation of the real conditions during observations from the assumed atmospheric model will result in a wrong reconstruction of the primary gamma-ray energy and the resulting source spectra.
During eight years of observations, the High Energy Stereoscopic System (H.E.S.S.) has experienced periodic natural as well as anthropogenic variations of the atmospheric transparency due to aerosols created by biomass burning.

In order to identify data that have been taken under such long-term
reductions in atmospheric transparency, a new monitoring quantity, the Cherenkov transparency coefficient, has been developed and will be presented here.
This quantity is independent of hardware changes in the detector and, therefore, isolates atmospheric factors that can impact the performance of the instrument, and in particular the spectral results.
Its positive correlation with independent measurements of the atmospheric optical depth (AOD) retrieved from data of the Multi-angle Imaging SpectroRadiometer (MISR) on board of the Terra NASA's satellite is also presented here.
}


\begin{document}
\maketitle

\section{Introduction}

During the last three decades, imaging atmospheric Cherenkov telescopes (IACTs) have qualified as powerful instruments for gamma-ray astronomy in the very-high-energy (VHE; E$>$0.1 TeV) regime, allowing detailed studies of the most violent phenomena known in the Universe.\\
The gamma-ray flux at these energies is rather low and the IACT technique provides the large effective areas required making use of telescopes on the ground.
Due to its opacity at these energies, the Earth's atmosphere acts as the calorimeter of the detector system, so therefore the VHE photons can be observed only indirectly at ground level. 
One of the main strengths of this type of detector is its low energy threshold, which unfortunately increases with the atmospheric absorption.\\
The atmospheric absorption will also affect the reconstruction of the energy of the primary particle, since shower images are compared to Monte Carlo shower simulations for which nominal hardware parameters and average atmospheric conditions at the H.E.S.S. site (23$^\circ$16'18'' S, 16$^\circ$30'00'' E, 1800\,m a.s.l) are assumed~\cite{Aharonian2006}~\cite{Bernlohr2000}. 
However this comparison might be affected by changes in the telescope efficiency, which includes not only changes in the optical efficiency and photo-sensor response but also atmospheric fluctuations. 
Any atmospheric phenomenon that acts as an atmospheric light absorber will attenuate Cherenkov light from EAS (Extensive Air Showers) particles and therefore reduce the amount of Cherenkov photons that reach the detector, which will cause an underestimation of energy of the primary gamma-ray.
This is especially problematic for spectral analysis since mis-reconstructed energies lead to biased values of the flux normalization and, in particular, in the case of non-power law spectra, other spectral parameters.\\
In this contribution we want to present a new way to estimate the atmospheric transparency by using only observables and calibration parameters from the Cherenkov data taken with the H.E.S.S. telescope array. We will introduce the most important atmospheric conditions that affect spectral shower reconstruction and a detailed comparison of this new atmospheric monitoring quantity with MISR (Multi-angle Imaging SpectroRadiometer) satellite data. Finally, the last part will contain a short systematic study on the effect of the atmospheric transparency on reconstructed spectral parameters.\\

\section{Atmospheric phenomenon}
Several atmospheric light absorbers attenuate the Cherenkov light emitted by the EAS, resulting in fewer photons reaching the camera and a lower trigger probability.
If absorbing structures (local clouds) are passing through the field of view, a fluctuating behavior in the central trigger rate~\cite{Funk2004} on time-scales smaller than the standard duration of data sets (28 min.) can be observed. Those data sets are removed from the final analysis sample through data quality cuts~\cite{Hahn2013}.

Thin layers of clouds and aerosol particles of human or natural origin might absorb or scatter photons from Cherenkov showers.
Even though the atmosphere at the H.E.S.S. site features a very low aerosol concentration, seasonal biomass burning to the north-east of the H.E.S.S. site around August and October may lead to a significant increase in the atmospheric aerosol content. These aerosols can be transported over large distances to the H.E.S.S. site and may persist in the atmosphere for weeks, resulting in an increase of the aerosol concentration over several kilometres in height at the site~\cite{Bernlohr2000}.

\section{Cherenkov transparency coefficient}

By monitoring only the trigger rates one is not able to dissentangle between the data taken under the presence of elevated aerosol concentrations and of large-scale clouds from the instrumental changes, specially during maintenance works.
Therefore we have developed a new quantity, the \textit{Cherenkov transparency coefficient}, designed to be as hardware-independent as possible in order to separate hardware-related effects from the decrease in trigger rates caused by large-scale atmospheric absorption.\\

For the definition of the Cherenkov transparency coefficient we assume that the zenith-corrected telescope trigger rates $R$ are dominated by cosmic-ray(CR) protons, with flux $f(E) = 0.096\cdot (E/\text{TeV})^{-2.70} \text{m}^{-2} \text{s}^{-1} \text{TeV}^{-1}\text{sr}^{-1}$ ~\cite{BESS98} at VHE.\\
Therefore $R\simeq k\cdot E_0^{-1.7 + \Delta}$, where $\Delta$ reflects the spectral shape of the effective area and $E_0$ is the energy threshold of the telescopes and is inversely proportional to the average pixel gain $g$~\cite{Aharonian2004}, the muon efficiency $\mu$~\cite{HESSMuons} and the atmospheric transparency, parametrized by a factor $\eta$ so that $ E_0 \propto (\eta \cdot \mu \cdot g )^{-1} $.
Therefore, for each telescope ``i'', $$\eta \propto \frac{R_i^{\frac{1}{1.7-\Delta}}}{\mu_i\cdot g_i} \equiv t_i.$$

Random fluctuations in the trigger of a single telescope are removed by selecting only the read-out events for which at least two telescopes are triggered in coincidence. This read-out trigger will therefore depend on the number of active telescopes so the averaged trigger over all N active telescopes is calculated and rescaled by a factor of $k_N$ that depends on the telescope multiplicity. The normalization to the peak position is yielded assuming $k_3 = 3.11$ and $k_4 = 3.41$ and will cancel the contribution of other CR species to the trigger rate.

\begin{figure}[h!!!]
  \includegraphics[width=0.5\textwidth]{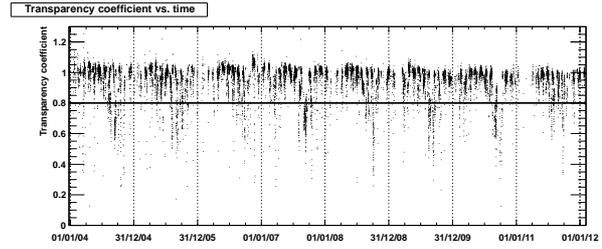}
  \caption{Evolution of transparency coefficients over the last 8 years. The solid line indicates the current data quality cut value at 0.8~\cite{Hahn2013}. The distribution is sharply peaked at 1 with a FWHM $\sim$ 9$\%$.}
  \label{figure01}
\end{figure}

The {\it Cherenkov transparency coefficient} (T) is then defined as
$$T \equiv \frac{1}{N\cdot k_N}\sum_i{t_i}.$$
Fig.~\ref{figure01} illustrates that the atmospheric transparency measure T is indeed independent of hardware effects and stable over time during an 8 year period of data taken by H.E.S.S., with peak values around 1.

\section{Aerosol atmospheric absorption}

The sensitivity of the Cherenkov transparency coefficient to the concentration of aerosols in the atmosphere will be confirmed through a positive correlation with independent aerosol measurements, such as satellite data.
The influence of the atmospheric aerosols on transparency is quite complex and strongly depends on the detailed scattering and absorption properties of the different aerosol types (e.g. sulphate, dust, organic carbon, sea salt).
Many studies of the atmospheric absorption of aerosols have been carried out, not only for astronomical purposes but also for climate and atmospheric studies.
However, we are particularly interested in those related with increases of aerosol absorption due to biomass burning. Such biomass burning aerosols seem to decrease the amount of UV solar radiation reaching the surface by up to 50\%, with typical values in the range of $\sim$ 15-35\%~\cite{Kalashnikova2007}. 
 
In the following we will test for a correlation between the Cherenkov transparency coefficient and the Aerosol Optical Depth (AOD); or more specifically with the atmospheric transparency ($\propto \exp(-\mathrm{AOD})$).

The MISR (Multi-angle Imaging SpectroRadiometer) instrument on board NASA's Terra spacecraft, has a better grid spatial resolution (1.1 km in global mode)~\cite{Diner1988} with respect to other satellite instruments and has the capability of observing at different viewing angles, so that MISR can distinguish between different types of atmospheric particles (aerosols), different types of clouds and different land surfaces.\\
 
The processed (Level 3) AOD data used in this study have proven to be in better agreement with the ground-based Aerosol Robotic Network (AERONET) measurements~\cite{Tesfaye2011} than previous satellite measurements.
In particular, a detailed 10-year study of the aerosol climatology with MISR over South Africa, Namibia's neighbour country, has revealed that the northern part of South Africa seems to be rich in aerosol reservoirs and the aerosol concentration (based on optical depth) is 34\% higher than that in the southern part of the country~\cite{Tesfaye2011}. 

\subsection{Correlation between Cherenkov transparency coefficient and MISR data}
Tesfaye et al. (2011) have also found seasonal changes in the aerosol composition in South Africa. During summer and early winter in the southern hemisphere, the northern part of South Africa is dominated by a mixture of coarse-mode and accumulation-mode particles, which are a result of air mass transport from arid/semi-arid regions of the central parts of South Africa, Botswana and Namibia. In the time from August to October (winter and early summer) it is dominated by sub-micron particles. The most important sources of sub-micron particles are industrial and rural activities (including mines and biomass burning).

The periodic drops in the Cherenkov transparency coefficient for the H.E.S.S. site (see Fig.~\ref{figure01}) correlate with the seasonal increase of sub-micron particles due to, among others, biomass burning like in the neighbouring South Africa. This gives an indication of the main atmospheric phenomenon responsible for the reduced trigger rates of some H.E.S.S. observations, especially in early summer, and points to the Cherenkov transparency coefficient as a good data quality parameter to monitor the atmosphere transparency. 

Therefore, we expect a strong and positive correlation with the AOD measured by satellites. To do this we used AOD retrieved from MISR data and the Cherenkov transparency coefficient from H.E.S.S. data. Both data sets cover the same period of time between 2004 and 2011.
The processed (level 3) MISR AOD data at the H.E.S.S. site (with a grid spacial resolution of 0.5$^\circ$x0.5$^\circ$) at three different wavelengths (443 nm, 555 nm and 670 nm), from UV to red wavelengths were used. Note that the satellite only measures the AOD during daytime and depending on latitude, the satellite samples a fixed location every 2 to 9 days. The overlap of satellite measurements and H.E.S.S. data taking is therefore sparse within a time interval of overlap of 24 hours, reducing the available data set for the correlation study (only 2\% of the H.E.S.S. data can be used).

\begin{figure}
  \includegraphics[width=0.5\textwidth]{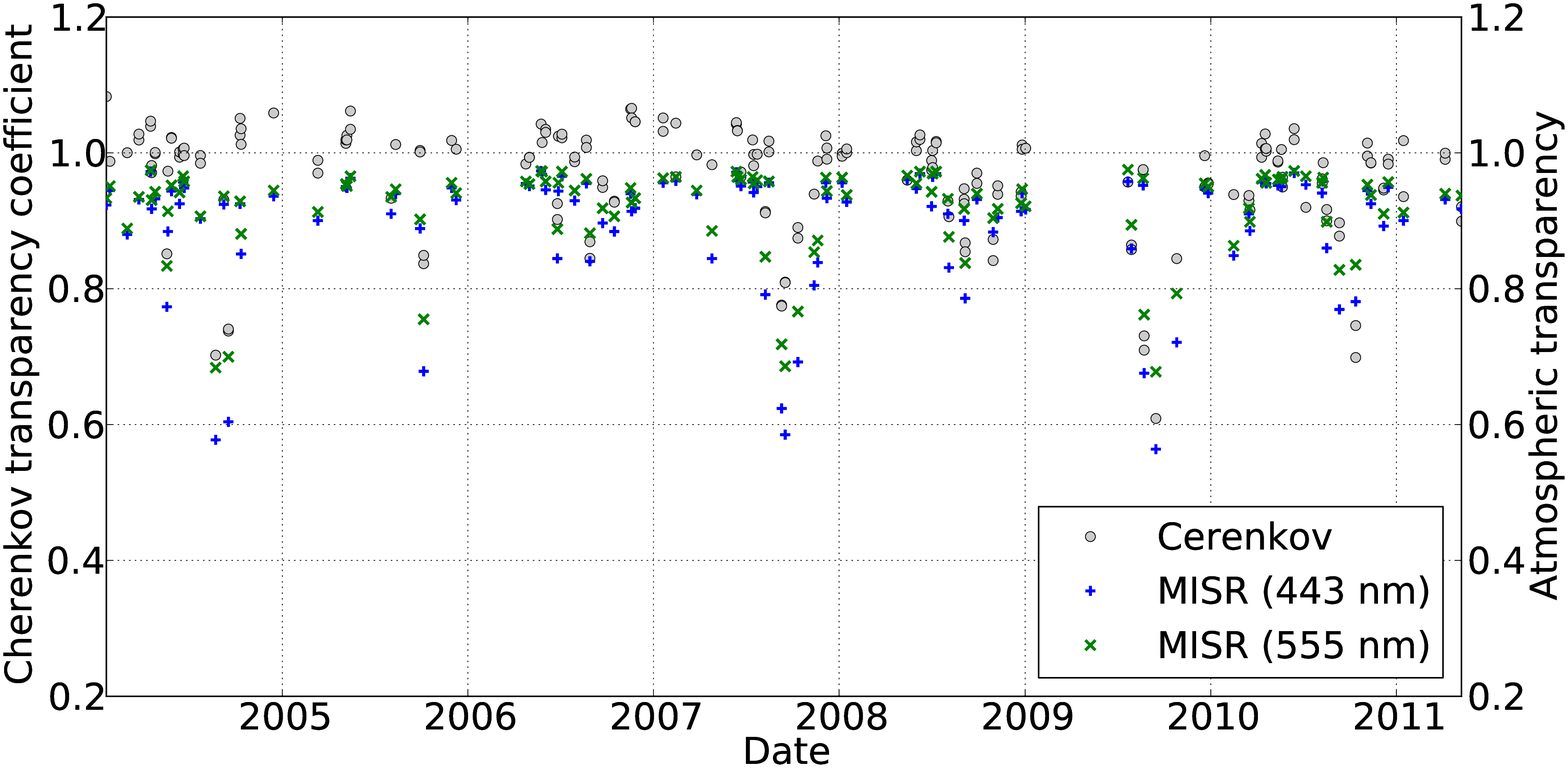}
  \caption{Cherenkov transparency coefficient measured in the time interval between 2004 and 2011, together with the MISR atmospheric transparency measurements in 443 nm (blue points) and 555 nm (green points).}
  \label{figure02}
\end{figure}

\begin{figure}
  \includegraphics[width=0.5\textwidth]{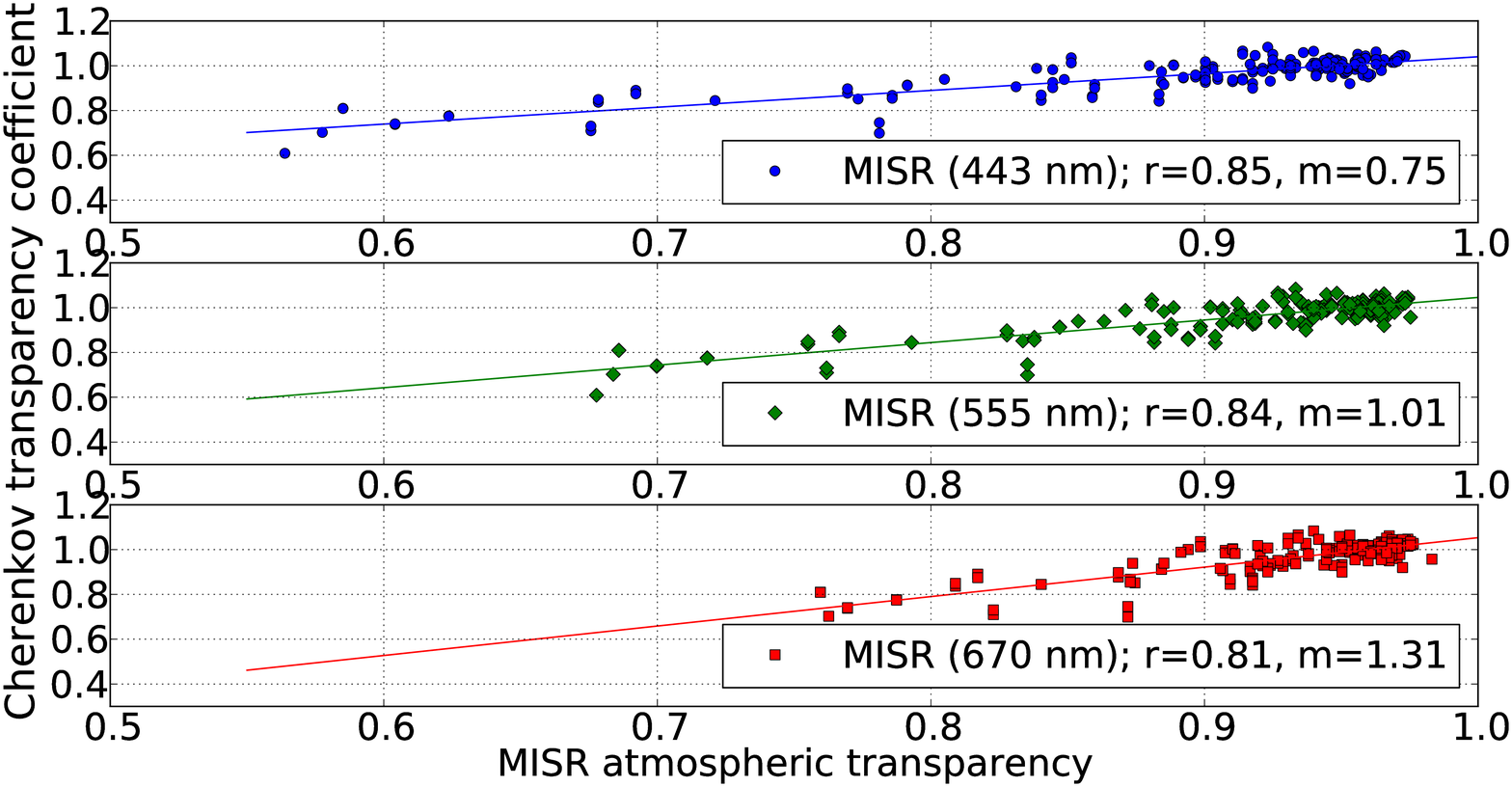}
  \caption{MISR atmospheric transparency ($\exp(-\mathrm{AOD})$) against the Cherenkov transparency coefficient. The three wavelengths measured by the MISR satellite are represented in different colors: 443 nm, blue; 555 nm, green and 670 nm, red. The resulting correlation is plotted as a solid line with the corresponding color of the MISR wavelength.}
  \label{figure03}
\end{figure}

Figure~\ref{figure03} quantifies the correlation between the atmospheric absorption ($\propto \exp(-\mathrm{AOD})$) for the three different wavelengths measured by MISR (443 nm in blue, 555 nm in green and 670 nm in red) and the Cherenkov transparency coefficient. The solid lines are the results of a linear fit between measurements ($T = m \cdot \exp(-\mathrm{AOD}) + c$).
The Pearson's correlation coefficients for the wavelengths 443 and 555 nm (blue and green) are $\sim$0.85 and $\sim$0.84 respectively. This shows a positive and strong correlation between the atmospheric transparency measured from satellites and the Cherenkov transparency coefficient, in particular for the blue band, which is most relevant for this study since the number of Cherenkov photons emitted per path length in a certain wavelength range (eq. (1) in ~\cite{Bernlohr2000}) is maximum in the UV-blue part of the spectrum.

Figure~\ref{figure03} also shows an increase of the steepness (``m'' in the figure) of the best fit of the linear correlation, with increasing wavelength. This is due to the fact that the atmospheric transparency measured with the MISR satellite decreases towards shorter wavelengths, while the Cherenkov transparency coefficient is always the same.

The decrease of the atmosphere transmission with decreasing wavelength can be explained by simple Mie scattering. Tesfaye et al. (2011) established an inverse proportionality between the aerosol particle size and their extinction efficiency at a certain wavelength. An increase of the AOD at short wavelengths therefore indicates the presence of sub-micron (radii$<$0.35 $\mu$m) particles, attributed by the authors to urban pollution (sulphates) and extensive biomass burning activities (carbonaceous aerosols). 

As a consequence, the aerosol-induced reduction in the atmosphere transparency is expected to be more pronounced at shorter wavelengths, which is where the bulk of the Cherenkov light is emitted.

\section{Systematic effects on reconstructed spectra}

For the study of the systematic effect caused by the different atmospheric transparencies on the spectral parameters we have analyzed the Crab Nebula data taken over 8 consecutive years from 2004 to 2011. During this period no detectable variability on flux has been detected from Crab Nebula at VHE energies.
The full data set investigated has an exposure of 84 hours, using only observations within one degree offset from the source. Also, to minimize a possible zenith-dependent energy bias, only data taken at zenith angles smaller than 47 degrees have been selected. 
The full data set has been divided into subsets of data corresponding to different ranges of the atmospheric transparency parameter after applying standard quality criteria to remove those runs with technical problems or with small clouds in the field of view during the observations~\cite{Aharonian2006}.
The standard cut-based analysis using simple air shower image parameters, the so called Hillas analysis ~\cite{Hillas1985}, was then employed to obtain spectral information for each subset.

The gamma ray spectrum of the Crab nebula has been measured by H.E.S.S. ~\cite{Aharonian2006} and was found to have an approximate power-law shape with some bending at the highest energies. For a pure power-law fit in the energy range ($0.41$-$40$)TeV, the flux normalization at 1 TeV, $\phi_0$, and the spectral index, $\Gamma$, were found to be $\phi_0=(3.45\pm0.05_{stat}\pm0.69_{sys})\times 10^{-11}\text{cm}^{-2}\text{s}^{-1}\text{TeV}^{-1}$ and $\Gamma=2.63\pm0.01_{stat}\pm0.10_{sys}$.

Assuming that atmospheric absorption leads to an underestimation of the reconstructed energy by a constant attenuation factor, a power law spectrum is expected to stay form invariant with changing atmospheric conditions while the flux normalization is expected to change. Quantitatively, assuming the reconstructed gamma ray energy $E_{reco}$ and the true energy $E_{true}$ to be related via $E_{reco}\propto T\cdot E_{true}$, 
one finds
\begin{eqnarray} \frac{\text{d}F}{\text{d}E_{true}} \propto E_{true}^{-\Gamma} ~~~~~\Leftrightarrow ~~~~~ \frac{\text{d}F}{\text{d}E_{reco}} \propto E_{reco}^{-\Gamma}\cdot T^{\Gamma-1}
\label{eqfit}
\end{eqnarray}

\begin{figure}[h!!!]
  \begin{center}
      \includegraphics[width=.3\textwidth]{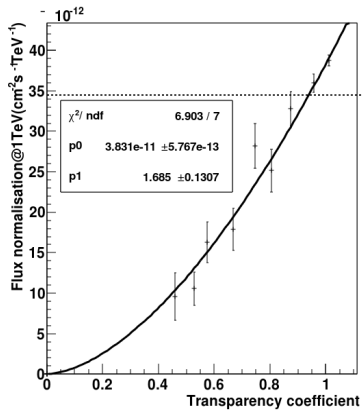}
  \end{center}
  \caption{Flux normalization at 1 TeV for Crab data taken during 8 years of H.E.S.S. operation. Abscissa values are given by the mean value of the transparency coefficient in the respective subset. The dashed line represents the published flux normalization value~\cite{Aharonian2006}.}
  \label{figure04}
\end{figure}

Fig.~\ref{figure04} displays the reconstructed flux normalization $\phi_0$ as function of the measured transparency coefficient $T$, confirming the expected strong T-dependence. Fitting a power law function $\propto T^{\Gamma-1}$ as expected from (eq.~\ref{eqfit}) yields a good description of the data (solid line in Fig. \ref{figure04}) with $\Gamma=2.69\pm 0.13$, in perfect agreement with the spectral index obtained from the power law fit to the Crab spectrum.
No significant T-dependence of the reconstructed spectral index $\Gamma$ has been found~\cite{Hahn2013}. Given the limited statistics, we conclude that the simple attenuation model (eq.~\ref{eqfit}) is at least a good approximation.

\section{Conclusions}

Changes in the central trigger rates of the H.E.S.S. array are caused both by the changes in telescope properties and in atmospheric conditions.
However, disentangling atmospheric extinction effects that take place on longer time scales from long-term trigger rate decreases connected to the aging of the instrument is an issue, specially during upgrades and maintenance work.
To deal with these challenges, a hardware-independent data quality quantity, the \textit{Cherenkov transparency coefficient}, has been developed to measure the atmospheric transparency, being sensitive to an increase in absorber concentrations.\\
Systematic studies reveal a bias of less than 20\% in the flux normalization applying data quality cuts sensible to the presence of atmospheric light absorbers (e.g. clouds and/or aerosols).
A large correlation factor, of about 0.85 at blue wavelengths ($\lambda$=443 nm), proves the relation between AOD measurements performed by the MISR satellite and the Cherenkov transparency coefficient, extracted directly from the Cherenkov technique.\\
However, several factors might limit the correlation, like the time difference between the satellite and H.E.S.S. measurements and the low statistics at low T due to low trigger rates. Also, the Cherenkov coefficients are not able to disentangle between large-scale clouds and layers of aerosols and are based on some simplified assumptions, such as the perfect inverse proportionality between telescope energy threshold and the muon efficiency. Addressing these points might result in a better correlation between the atmospheric absorption measured by satellites and the Cherenkov transparency coefficient.
Simultaneous observations of on-site Radiometer and LIDAR data and the Cherenkov telescope might also help and are currently under study~\cite{RadiometerLIDAR}.\\
The Cherenkov transparency coefficient is currently used as a data quality parameter in H.E.S.S.. Previous methods, using other atmosphere-sensitive parameters~\cite{HEGRA},~\cite{VERITAS},~\cite{MAGIC} and~\cite{Chadwick}, have been used to correct the flux for changes in atmospheric conditions.
The strong correlation with independent atmospheric measurements suggests that the Cherenkov transparency coefficient could be applied in the same way, currently under study, making it possible to use the Cherenkov technique over a wider range of atmospheric conditions.

\vspace*{0.5cm}
\footnotesize{{\bf Acknowledgment:}{
The authors would like to acknowledge the support of their host institutions. We want to thank the whole H.E.S.S. collaboration for their support, especially Prof. Werner Hofmann and Prof. Christian Stegmann as well as Prof. Thomas Lohse and Dr. Ira Jung for the many fruitful discussions.
}}


\begin{thebibliography}{}
\bibitem{Aharonian2004} Aharonian, F. et al., Astropart.Phys. 22 (2004) 109-125.
\bibitem{Aharonian2006} Aharonian, F. et al.,A$\&$A 457 (2006) 899-915.
\bibitem{Bernlohr2000} Bernl\"ohr, K. , Astropart.Phys. 12 (2000) 255-268.
\bibitem{RadiometerLIDAR} Chaves et al., 2013, these proceedings (ID 1122).
\bibitem{Diner1988} Diner, D.J. et al, IEEE Trans. Geosci. Rem. Sens.36 (1988) 1072-1087. 
\bibitem{HESSMuons} Bolz, PhD thesis (2004).
\bibitem{MAGIC} Dorner, D. et al, A\&A 493 (2009)721.
\bibitem{Funk2004} Funk, S. et al., Astropart.Phys. 22 (2004) 285-296.
\bibitem{Hahn2013} Hahn J. et al, 2013, submitted to Astropart.Phys.
\bibitem{Hillas1985} Hillas, A.~M., International Cosmic Ray Conference 1985 3 (1985) 445-448.
\bibitem{Kalashnikova2007} Kalashnikova, O.V., et al, Rem. Sens. Environ. 107 (2007) 65-80.
\bibitem{VERITAS} LeBohec, S.; Holder, J., Astropart. Phys. 19 (2003) 221-233.
\bibitem{Chadwick} Nolan, S.J. et al, Astropart.Phys.  34 (2010) 304.
\bibitem{HEGRA} P\"uhlhofer, G. et al, Astropart.Phys. 20 (2003) 26.
\bibitem{BESS98} Sanuki, T. et al, ApJ 545 (2000) 1135.
\bibitem{Tesfaye2011} Tesfaye, M. et al, Journal Of Geophysical Research 116 (2011) 20216.



\end{thebibliography}
\end{document}